\documentclass[]{aa}
\usepackage{graphicx}                         

\begin{document}

\title{The distribution of the ISM in the Milky Way}
\subtitle{A three-dimensional large-scale model}

\author{A. Misiriotis \inst{1}
\and E. M. Xilouris \inst{2}
\and J. Papamastorakis \inst{1}
\and P. Boumis \inst{2}
\and C. D. Goudis \inst{2,3}}

\offprints{Angelos Misiriotis, \email{angmis@physics.uoc.gr}}

\institute{University of Crete, Physics Department, P.O. Box 2208, 710 03
Heraklion, Crete, Greece
\and National Observatory of Athens, I. Metaxa \& Vas. Pavlou Str., 
Palaia Penteli, GR-15236, Athens, Greece
\and Astronomical Laboratory, Department of Physics, University of Patras, GR-26110 Patras, Greece}
\date{Received / Accepted}
\abstract{We use the COBE/DIRBE (1.2, 2.2, 60, 100, 140, and 240 $\mu$m) maps
and the COBE/FIRAS spectra (for the wavelength range 100 - 1000 $\mu$m) to constrain
a model for the spatial distribution of the dust, the stars, and the gas in the Milky Way.
By assuming exponential axisymmetric distributions for the dust and the stars and by
performing the corresponding radiative transfer calculations we closely 
(given the simple geometry of the model) reproduce the FIR and NIR maps of the Milky Way.
Similar distributions for the atomic and molecular hydrogen in the disk are used 
(with an inner cut-off radius for the atomic hydrogen) to fit the gas data. 
The star formation rate as a function of the Galactic radius is derived from the FIR emission
and is well in agreement with existing estimates from various star formation tracers.
The gas surface density is plotted against the star formation rate density and an ``intrinsic''
Galactic Schmidt law is derived with excellent agreement with the ``external''
Schmidt law found for spiral galaxies. 
The Milky Way is found to consume $\sim 1\%$ and $\sim 10\%$ of its gas 
in the outer and inner regions respectively (for a period of 0.1 Gyr) to make stars.
The dust-induced B-V color excess observed in various directions and distances (up to $\sim 6.5$ kpc)
with well-studied Cepheid stars is compared with the model predictions showing a good agreement. 
The simple assumption of exponential distributions of stars and dust in the Galaxy is found to 
be quite instructive and adequate in modeling all the available data sets from 0.45 $\mu$m (B-band) 
to 1000 $\mu$m.

}

\maketitle

\section{Introduction}
Our Galaxy, the Milky Way, constitutes the best laboratory for studying the
properties of the interstellar medium (ISM) in spiral galaxies. 
Being inside the Galaxy, the observer has the advantage
of looking through different lines of sight passing both through dense 
environments (e.g., the Galactic center) and through areas almost free of dust and gas
(e.g., the Galactic poles). The way that dust and gas (both in molecular and neutral phases) are
distributed inside the Galaxy not only shapes the appearance of the Galaxy
in different wavelengths, but also drives star formation, a key process that
determines Galactic evolution.

There are several tracers for measuring the star formation 
activity in a galaxy (e.g., ultra-violet (UV) emission, far-infrared (FIR) emission, 
{\rm H$_\alpha$}, and radio emission), yet it is far from trivial to get an accurate estimate
(see Kennicutt, \cite{kennicutt_1998b} for a review). 
Most of the star formation rate (SFR) tracers are affected by dust extinction, thus rendering 
their use quite limited and uncertain (see, however, Kewley et al. \cite{kewley} for 
a discussion on properly accounting for extinction).
The far infrared (FIR) emission, if modeled properly to distinguish 
between the diffuse dust emission and the emission coming from the
star formation complexes, has been proven to be one of the most promising SFR indicators 
(see Misiriotis et al. \cite{misiriotis_2004}, and for a review Kylafis \& Misiriotis \cite{kylafis2}).

Some efforts have been made in modeling the interstellar medium (ISM) in the Milky Way. 
Kent et al. (\cite{kent}) used the K-band data from the SPACELAB infrared telescope
and modeled the distribution of the light in the Galaxy by 
assuming an exponential distribution for the starlight in the disk and the bulge.
Using the COBE 140 and 240 $\mu$m maps along with radio surveys of HI and H$_2$,
Sodroski et al. (\cite{sodroski}) modeled the FIR emission of the Galaxy and derived
the dust temperature profiles as well as gas-to-dust mass ratios and emissivities
along different lines of sight. 
A composite of COBE/DIRBE and IRAS/ISSA maps made by Schlegel et al. (\cite{schlegel}) 
provides a unique tool for estimating the Galactic extinction along different directions.
Davies et al. (\cite{davies}) modeled the FIR emission from the Galaxy 
by fitting exponential dust distributions
to the COBE/DIRBE maps. In this study, the authors find that the best fit could be
achieved by assuming that the dust distribution is more extended than that of the stars,
namely that the scalelength of the dust is 1.5 times that of the stars and the scaleheight
of the dust is twice that of the stars. Modeling of nearby edge-on galaxies (e.g.,
Xilouris et al. (\cite{xilouris3}) and references therein) 
do support the finding of the dust having a larger scalelength with respect to the stellar one, 
but not the scaleheight relation found in this study. In our analysis we will perform 
modeling of the near-infrared (NIR) and the FIR maps of the Milky Way and thus we will
investigate the relation between the dust and stellar geometrical characteristics.
Drimmel \& Spergel (\cite{drimel1}) presented a very detailed modeling
of the Milky Way dust content. Based on a total of 48 parameters (26 parameters for the dust 
and 22 parameters for the stars), this model
quite nicely fitted the NIR (J-band) and FIR (240 $\mu$m) radial profiles.
In a subsequent paper (Drimmel et al. \cite{drimel2}), this model was used to calculate
the extinction to any point within the Galactic disk.
This model provides a detailed description of the Galactic morphology,
taking into account the structures of the disk, the spiral arms, and the warp.
In contrast, our approach aims to model the Galaxy using as few parameters 
as possible. We know in advance that we will not be able to get an
excellent representation of the detailed morphology, but we will be able to
 draw results on the large-scale distribution of the dust, the
stars, and the star formation in our Galaxy.

In this paper we use a simple (with the minimal number of parameters
required), but realistic, three-dimensional model, which includes the effects of
absorption and scattering of the stellar light by the dust in the optical and near infrared 
(NIR) wavelengths. We also take into account the emission from the diffuse dust  
and the emission from a warm dust component associated with the star-forming regions.
In order to constrain the model parameters, we fit the model to all the
available data simultaneously. The data that we use consist of the COBE/DIRBE 
(1.2, 2.2, 60, 100, 140, and 240 $\mu$m) maps and the COBE/FIRAS spectra 
(for the wavelength coverage between 100 and 10000 $\mu$m). 
Using published maps for the atomic and molecular hydrogen, we model the 
gas distribution. After calculating the SFR density throughout the Galactic disk 
and comparing it with the observed gas surface density, we investigate the validity of the 
Schmidt (\cite{schmidt}) law as quantified by Kennicutt (\cite {kennicutt_1998a}). 
Finally, using a catalog of Cepheid stars with good estimates of their distance and B-V color 
excess, we confirm the validity of our model in the optical region.

\section{The data}
\subsection{The COBE data}
The Cosmic Background Explorer (COBE) satellite, 
launched on November 18, 1989, has been proved to be a very
successful space mission (Boggess et al. \cite{boggess}).
With the Diffuse Infrared Background Experiment (DIRBE) instrument, COBE
surveyed the entire sky in 10 photometric bands covering the wavelength region
from 1 to 240 $\mu$m at an angular resolution of 0.7 square degrees (Sodroski et al. \cite{sodroski}).
Using another instrument, the Far Infrared Absolute Spectrometer (FIRAS),
COBE measured the spectrum of the dust emission of our Galaxy in the wavelength
range from 120 $\mu$m to 10000 $\mu$m (Fixsen et al. \cite{fixsen}). 
These two instruments provide us with a unique set of data of the dust
emission for a wide range of wavelengths, ideal for studying the properties and
the distribution of dust in the Galaxy.

\subsubsection{NIR and FIR maps}
DIRBE was designed primarily to conduct a systematic search for an 
isotropic cosmic infrared background in 10 photometric bands from
1.25 to 240 $\mu$m (Boggess et al. \cite{boggess}; Hauser et al. \cite{hauser}). 
Boggess et al. (\cite{boggess}) report an rms sensitivity per field of view (0.7 square degrees)
of 1, 0.9, 0.4, 0.1, 11, and 4 ($\rm{10^9~W m^{-2} sr^{-1}}$) at
1.2, 2.2, 60, 100, 140, and 240 $\mu$m, respectively.

For the purposes of this study we make use of the 1.2, 2.2, 60, 100, 140, and 240 $\mu$m 
Zodi-Subtracted Mission Average maps obtained from the COBE database. 
This set of data consists of weekly averaged intensity maps with the zodiac
light subtracted from the maps (see Kelsall et al. \cite{kelsall} for a description
of the model for the interplanetary dust).    
The data were extracted from the original data-cubes and analyzed using 
the special COBE analysis software 
UIDL\footnote{http://lambda.gsfc.nasa.gov/product/cobe/cgis.cfm} written in IDL.

\subsubsection{FIR and submillimeter spectra}
FIRAS is a polarizing Michelson interferometer (Mather \cite{mather}) with two separate spectral
channels (the high-frequency channel, extending from 120 to 500 $\mu$m and the low-frequency
channel extending from 500 to 10000 $\mu$m). The in-orbit absolute calibration of
FIRAS was accomplished by using an external black-body calibrator. The spectra, already
corrected for Cosmic Microwave Background and zodiac light contribution, were extracted
from the original data-cubes and averaged for eight different directions
on the sky over a range of 20 degrees in Galactic latitude (from -10 to 10 degrees)
and from 0 to 10, 5 to 15, 15 to 25, 25 to 35, 55 to 65, 85 to 95, 115 to 125, and
135 to 145 degrees in Galactic longitude.

\subsection{The {\rm HI and H$_2$} maps}
The Leiden/Argentine/Bonn survey of the Galactic HI (Kalberla et al. \cite{kalberla}) is a 
merging of the Argentino de Radioastronom\'{i}a and the Leiden/Dwingeloo surveys.
The angular resolution of the combined survey is 0.6 square degrees (comparable to that of the 
COBE maps) and is the most sensitive, to date, HI map of the Milky Way.

A composite CO survey of the Milky Way constructed from 37 individual surveys has been 
carried out by Dame et al. (2001)\footnote{http://cfa-www.harvard.edu/mmw/MilkyWayinMolClouds.html}.
The total area covered by this survey is 9,353 square degrees, which accounts
for nearly one half of the area within 30 degrees of the Galactic plane.
Using a CO-to-H$_2$ mass conversion factor of
$X=1.8\pm0.3 \times 10^{20}$ cm$^{-2}$ K$^{-1}$ km$^{-1}$ s (Dame et al. \cite{dame}),
a H$_2$ map is constructed.

\section{The model}
\label{sec:model}
The model that we use quite accurately simulates the physical processes 
(see Kylafis \& Bahcall \cite{kylafis1} and Popescu et al. \cite{popescu1})
that take place inside a galaxy for the wavelength range between optical and
submillimeter (submm).
In the optical/NIR wavelength range, absorption and scattering of the stellar light
by the dust grains occurs, while emission from the heated dust grains is taking
place in the FIR/submm wavelengths. The different nature of the processes allows
us to treat the two wavelength ranges separately as we describe below. Since 
the dust reveals itself by its FIR/submm emission, it is reasonable to begin
with the modeling in these wavelengths to constrain the dust parameters.
By fixing the dust parameters in the radiative transfer model, we can then model
the optical/NIR emission of the Galaxy.

For computational reasons, all the radiative transfer calculations
are made inside a cylinder of a radius of three times the scalelength of the 
dust and a half-height of six times the scaleheight of the stars.

\subsection{The FIR/submm wavelength range}

The emission model that is used for the FIR/submm wavelengths is along the
lines of the model described in Popescu et al. (\cite{popescu1}; see also Misiriotis
et al. \cite{misiriotis_2001}). Following these studies, 
we assume that the Galactic dust can be described by two components.
The first component is the warm dust associated with star formation complexes, which
is heated locally by the young stars.
The second component is the cold diffuse dust, which is heated by the diffuse radiation 
field in the Galaxy.
For the dust grain emissivity in these wavelengths, we use the values reported
in Weingartner \& Draine (\cite{weingartner}).

\subsubsection{The warm dust component}
For the density of the warm dust component we assume that its three-dimensional distribution
is given by the following equation 
\begin{equation}
\rho_{w}(R,z)=\rho_{w}(0,0)\exp \left( - \frac{R}{h_w} - \frac{|z|}{z_w} \right),
\end{equation}
where $R$ and $z$ are the cylindrical coordinates, $\rho_{w}(0,0)$ is the
warm dust density at the center of the Galaxy and $h_w$ and $z_w$ are the exponential
scalelength and scaleheight respectively. 
The emission of the dust depends on the temperature $T_w$ and on the emission
cross-section per unit mass of dust $\kappa_\lambda$. 
In the wavelengths longwards of 60$mum$, the extinction and the emission cross-sections
per unit mass have the same numerical value because the albedo is negligible.
Therefore, throughout the paper we will use the same symbol $\kappa_\lambda$ to denote 
the emission and extinction cross-sections of the dust.
For the temperature of the warm dust $T_w$, we assume that
it is constant throughout the Galaxy and equal to 35 K.
This is a reasonable assumption since the warm dust is mostly associated with
regions of intense star formation. The spectral energy distribution (SED)
of a typical star-forming region can
be approximated by thermal emission from dust with a temperature of 35 K 
(see Popescu et al. \cite{popescu1} and Misiriotis et al. \cite{misiriotis_2004}).
For $\kappa_\lambda$ we use the $R_V = 3.1$ model from Weingartner \& Draine (\cite{weingartner}). 
Thus, at a given wavelength $\lambda$, at a given point $(R,z)$ in the Galaxy, 
the emission per unit volume per unit wavelength per steradian $\eta_w(\lambda,R,z)$
can be calculated by
\begin{equation}
\eta_w(\lambda,R,z) = \rho_{w}(R,z) \kappa_\lambda B(\lambda,T_w),
\label{eq:warmdustemission}
\end{equation}
where  $B(\lambda,T)$ is the Planck function.

\subsubsection{The cold dust component}
The density of the cold dust component is described by a similar exponential disk 
according to the following equation
\begin{equation}
\rho_{c}(R,z)=\rho_{c}(0,0)\exp \left( - \frac{R}{h_c} - \frac{|z|}{z_c} \right).
\end{equation}
Here $\rho_{c}(0,0)$ is the cold dust density at the center of the Galaxy and $h_c$ and $z_c$ are
the exponential scalelength and scaleheight, respectively. For the temperature of the cold dust,
we assume that it follows an exponential distribution within the Galactic disk, namely,
\begin{equation}
T_c(R,z)=[T_c(0,0)-T_\infty] \exp \left( -\frac{R}{h_T} - \frac{|z|}{z_T} \right) + T_\infty.
\label{eq:colddusttemp}
\end{equation}
Here $h_T$ and $z_T$ are the exponential scalelength and scaleheight for the temperature
of the cold dust with $T_c(0,0)$ being the temperature at the center of the Galaxy and $T_\infty$
the temperature of the dust at infinity (in our case taken equal to 3 K).
The emission of the cold dust per unit volume per unit wavelength per steradian $\eta_w(\lambda,R,z)$
is given by the formula

\begin{equation}
\eta_c(\lambda,R,z) = \rho_{c}(R,z) \kappa_{\lambda} B(\lambda,T_c(R,z)),
\label{eq:colddustemission}
\end{equation}
which is identical to Eq. (\ref{eq:warmdustemission}) for the warm dust component with the 
dust temperature now varying within the galaxy according to Eq. (4).

Given Eqs.(\ref{eq:warmdustemission}) and (\ref{eq:colddustemission}), the FIR/submm intensity
along any line of sight $I(l,b,\lambda)$ can be calculated by
\begin{equation}
I(l,b,\lambda)=\int_{s_0}^{Earth} ds [\eta_w(\lambda,R,z) + \eta_c(\lambda,R,z)],
\label{eq:totdustemission}
\end{equation}
where $l$ and $b$ are the galactic longitude and latitude, respectively. The Earth is assumed to
lie exactly on the galactic plane and at a distance equal to 8 kpc from the galactic center (Bahcall \&
Soneira \cite{bahcall}). The lower limit of the integral $s_0$ is a point where the line of sight 
intersects a ``box'' inside which we assume that all the galaxy is contained.

In total, the dust spatial distribution is described by 10 parameters (namely, 
$\rho_{w}(0,0), h_w, z_w, \rho_{c}(0,0), h_c, z_c, T_c(0,0), h_T$, $z_T$, and $T_w$),
which can be constrained using the COBE/DIRBE data at 60, 100, 140, and 240 $\mu m$.

\subsection{The optical/NIR wavelength range}
Given the dust distribution, as this is inferred by the FIR emission, one can calculate the
extinction along any line of sight at any distance from the Earth. Using this information,
we calculate the observed starlight intensity along any line of sight at 1.2 and 2.2$\mu m$. 
In particular, the intensity is given by the following equation (Kylafis \& Bahcall \cite{kylafis1})

\begin{equation}
I(l,b,\lambda)=\int_{s_0}^{Earth}ds~\eta_s(\lambda,R,z) e^{-\int_{s}^{Earth}ds'\kappa_\lambda \rho_c(R,z)}.
\label{eq:niremission}
\end{equation}
Here $\eta_s(\lambda,R,z)$ is the stellar emissivity at wavelength $\lambda$ at position $(R, z)$ 
in the  galaxy, $\kappa_\lambda$ is the extinction cross-section per unit mass at wavelength $\lambda$ 
(taken from Weingartner \& Draine \cite{weingartner}), and 
$\rho(R, z)$ is the dust density. To simplify the calculations, we assume that only the diffuse 
cold dust component contributes to the opacity, while the warm dust component, being
associated mostly with the star-forming complexes (see Sect. \ref{sec:model}), affects the opacity
locally without affecting the global properties of the Galaxy. As we will also see
later (Sect. \ref{sec:results}), the mass of the warm dust component is more than 2 orders of
magnitude less than that of the cold dust component, making it only a very small fraction
of the ISM. For the spatial distribution of the stellar emissivity we follow Xilouris et al.
(\cite{xilouris1}, \cite{xilouris2}, \cite{xilouris3})
and assume that the stellar distribution is described by

\begin{displaymath}
\eta_s(R,z) = \eta_{disk} \exp \left( - \frac{R}{h_s} - \frac{|z|}{z_s} \right)
\end{displaymath}
\begin{equation}
~~~~~~~~~~+\eta_{bulge} \exp (-7.67 B^{1/4}) B^{-7/8}.
\end{equation}
In this expression the first part describes an exponential disk and
the second part describes the bulge, which in projection is the well-known
$R^{1/4}$-law (Christensen \cite{christensen}).
In the first term of the expression, $\eta_{disk}$ is the stellar emissivity
at the center of the disk, and $h_s$ and $z_s$ are the scalelength and 
scaleheight, respectively, of the stars in the disk.
For the bulge, $\eta_{bulge}$ is the stellar emissivity per unit volume
at the center of the bulge, while $B$ is defined by
\begin{equation}
B = \frac{\sqrt{R^2 + z^2 (a/b)^2}}{R_e} ,
\end{equation}
with $R_e$ being the effective radius of the bulge and $a$ and $b$ being the
semi-major and semi-minor axis, respectively, of the bulge.

To model the Galaxy in the NIR wavelengths (and in particular
at 1.2 and 2.2$\mu$m for which we have available full maps of the Galaxy),
we need to calculate 9 parameters. Three of them ($\rho_{c}(0,0), h_c, z_c$;
the parameters of the cold dust component) are constrained from the 60, 
100, 140, and 240 $\mu$m data, as we will see later on (see Sect. 4). The remaining 
6 parameters ($\eta_{disk}, h_s, z_s, \eta_{bulge}, R_e$, and $a/b$)
are constrained from the 1.2 $\mu$m map. Additional modeling is
also performed on the 2.2 $\mu$m map of the Galaxy (see Sect. \ref{sec:fit}). 

\subsection{The molecular and atomic hydrogen}
The two gas phases (${\rm HI}$ and ${\rm H_2}$), being totally independent from the dust and stars,
are treated separately from the rest of the components. 
The distribution of the molecular hydrogen (${\rm H_2}$) is assumed to follow the CO emission. 
Although it is rather clumpy, it can be, to a first approximation, described by an 
axisymmetric exponential disk (e.g., Regan et al. 2001 for the the large-scale
distribution of CO in other nearby galaxies). 
Thus, the distribution of ${\rm H_2}$ as a function of the position in the galaxy is given by
\begin{equation}
\rho_{\rm H_2}(R,z)=\rho_{\rm H_2}(0,0)\exp \left( - \frac{R}{h_{\rm H_2}} - \frac{|z|}{z_{\rm H_2}} \right),
\end{equation}
where $\rho_{\rm H_2}(0,0)$ is the density of the ${\rm H_2}$ molecules in the 
center of the galaxy, $h_{\rm H_2}$ is the scalelength, and $z_{\rm H_2}$ is 
the scaleheight of the distribution.

For the atomic hydrogen (HI), we assume a similar distribution, but we also introduce an inner
truncation radius to account for the absence of HI in the central parts of the Galaxy, 
as indicated by the relevant map (see Kalberla et al. \cite{kalberla}).
The distribution of HI is then given by
\begin{equation}
\rho_{\rm HI}(R,z)= 
\left\{
\begin{array}{lll}
\rho_{\rm HI}(0,0)\exp \left( - \frac{R}{h_{\rm HI}} - \frac{|z|}{z_{\rm HI}} \right) &, & \mbox{$\rho>R_{t}$}\\
0									  &, & \mbox{$\rho<R_{t}$} 
\end{array}\right.
\end{equation}
with $\rho_{\rm HI}(0,0)$ as the $HI$ density at the center of the galaxy (of the untruncated disk),
$R_{t}$ as the inner truncation radius, $h_{\rm HI}$ as the scalelength, and $z_{\rm HI}$ 
as the scaleheight of the distribution.
Although more sophisticated functional expressions have been suggested for the description of the 
HI distribution (i.e., van den Bosch \cite{vandenbosch}), we
adopt the simplest possible description.

The model column density of the molecular and the atomic hydrogen along any line of sight is 
then calculated by 
\begin{equation}
I_{\rm gas}(l,b)=\int_{s_0}^{Earth}ds~\rho_{\rm gas}(R,z)
\label{eq:gasemission}
\end{equation}
with $\rho_{\rm gas}(R,z)$ being equal to either the molecular $\rho_{\rm H_2}(R,z)$ or the atomic
$\rho_{\rm HI}(R,z)$ hydrogen distribution (see Eqs. (10) and (11), respectively).
The molecular hydrogen distribution is then parametrized by three parameters ($\rho_{\rm H_2}(0,0),
h_{\rm H_2}$, and $z_{\rm H_2}$), while the atomic hydrogen distribution, by four ($\rho_{\rm HI}(0,0),
h_{\rm HI}$, $z_{\rm HI}$, and $R_t$).

\section{The fitting procedure}
\label{sec:fit}
%%%%%%%%%%%%%%%%%%%%%%%%%%%%%% TABLE 1 %%%%%%%%%%%%%%%%%%%%%%%%%%%
\begin{table}
\centering
\caption[]{Masked sources in DIRBE maps.}
\begin{tabular}{lcc}
\hline
 Source & Central $l, b$ & Subtracted area  \\
        & (degrees) & (sq. degrees)  \\
\hline
Galactic center   &  0, 0       &  $10 \times 10$     \\
Orion arm         &  -80, 0     &  $20 \times 20$     \\
High latitude     &  10, 20     &  $20 \times 20$     \\
Magellanic clouds &  80, -30    &  $20 \times 20$     \\
M31               &  -120, -20  &  $30 \times 20$     \\
\hline
\end{tabular}
\label{tab:mask}
\end{table}
%%%%%%%%%%%%%%%%%%%%%%%%%%%%%%%%%%%%%%%%%%%%%%%%%%%%%%%%%%%%%%%

Having a computationally intensive radiative transfer model with 15 parameters in total to compute 
(9 for the dust distribution and 6 for the stellar distribution; see Sect. 3),
we proceed in two steps. We first constrain the 9 parameters that define the
dust distribution and the temperature of the cold dust component
by fitting the model FIR intensity (Eq. \ref{eq:totdustemission}) to the FIR data (i.e., 60, 100, 140, and 240$\mu$m).
Then, having defined the dust distribution, we fit Eq. (\ref{eq:niremission}) to the NIR maps
to constrain parameters that describe the stellar distribution.

A third step, treated separately, is to determine the parameters that describe the ${\rm H_2}$ and HI
distributions.

In all three steps, the observed surface brightness (or the column density in the case of the gas)
is compared with the computed surface brightness from the model.
Before going into a detailed $\chi^2$ minimization to find those values
of the parameters that best describe the Galaxy, we fit simple exponential profiles to the surface 
brightness and in directions vertical and parallel to the disk (see
Xilouris et al. 1997 for more details). In this way, good estimates of the
geometrical characteristics are derived and are used as initial guesses
in a $\chi^2$ minimization algorithm.
The minimization is done using the Steve Moshier C translation of the public
domain Levenberg-Marquardt solver of the Argonne National Laboratories MINPACK mathematical 
library\footnote{http://www.netlib.org}.
We always test the uniqueness of the best values of the parameters derived by
the fit by altering the initial values of the fit by as much as $\sim 30\% - 40\%$.
In all cases the values returned by the fit were the ones that we present.

For all the DIRBE maps, we used the data within the latitude range of -40 to 40 degrees.
This was done because outside this latitude range the signal drops significantly
with no usable input to the model. 
Some regions were also masked from the DIRBE maps due to their
large deviation from the Galactic emission (see Table~\ref{tab:mask}).

\section{Results}
\label{sec:results}

%%%%%%%%%%%%%%%%%%%%%%%%%% FIG 1 %%%%%%%%%%%%%%%%%%%%
\begin{figure*}
\resizebox{\hsize}{!}{\includegraphics{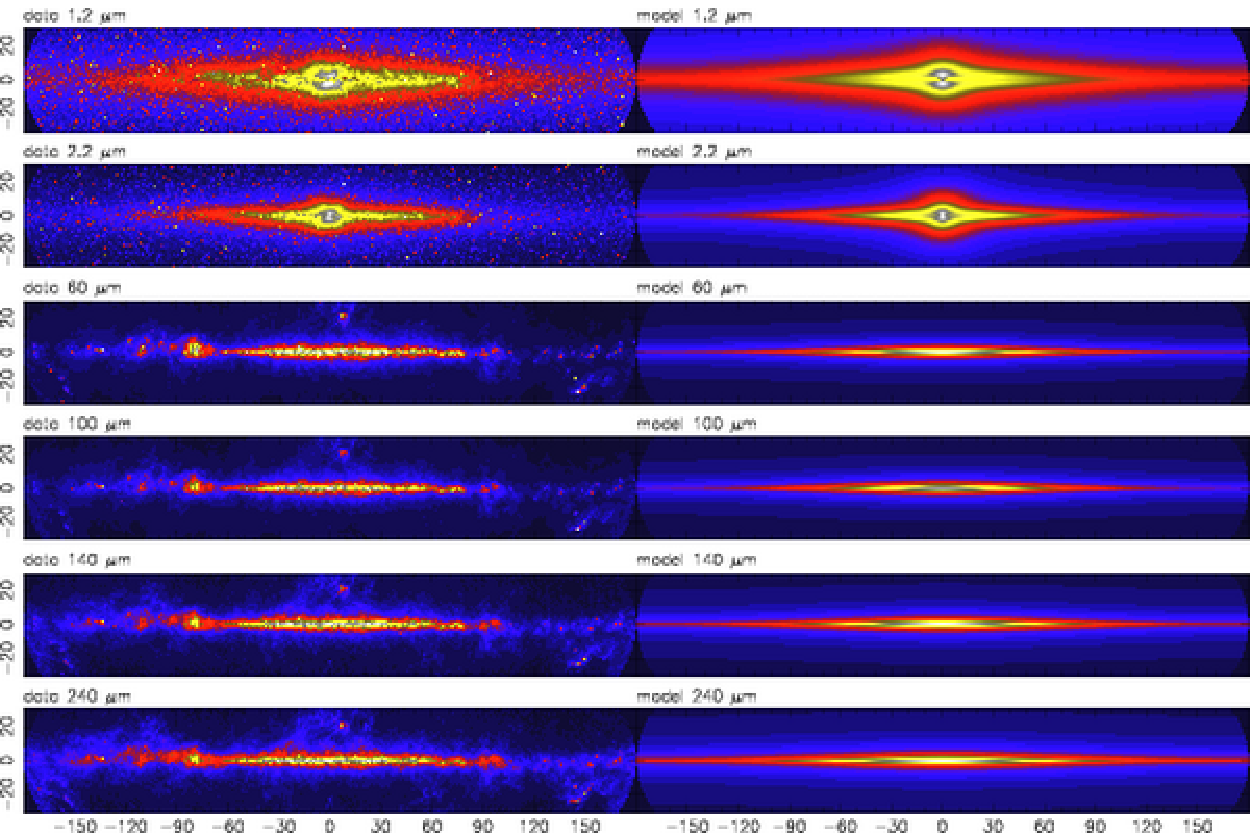}}
\caption
{
The COBE/DIRBE maps of the Galaxy (left panels) in direct comparison
with the fitted model (right panels). From top to bottom the
1.2, 2.2, 60, 100, 140, and 240$\mu$m maps are presented for the
Galactic latitude range of -30 to 30 degrees. The maps were made in Mollweide 
projection (Snyder \cite{snyder}).
}
\label{fig:images}
\end{figure*}
%%%%%%%%%%%%%%%%%%%%%%%%%%%%%%%%%%%%%%%%%%%%%%%%%%%%%%

%%%%%%%%%%%%%%%%%%%%%%%%%%%%%% FIG 2 %%%%%%%%%%%%%%%%%%%%%%%%%%%
\begin{figure*}
\includegraphics[width=17.6cm]{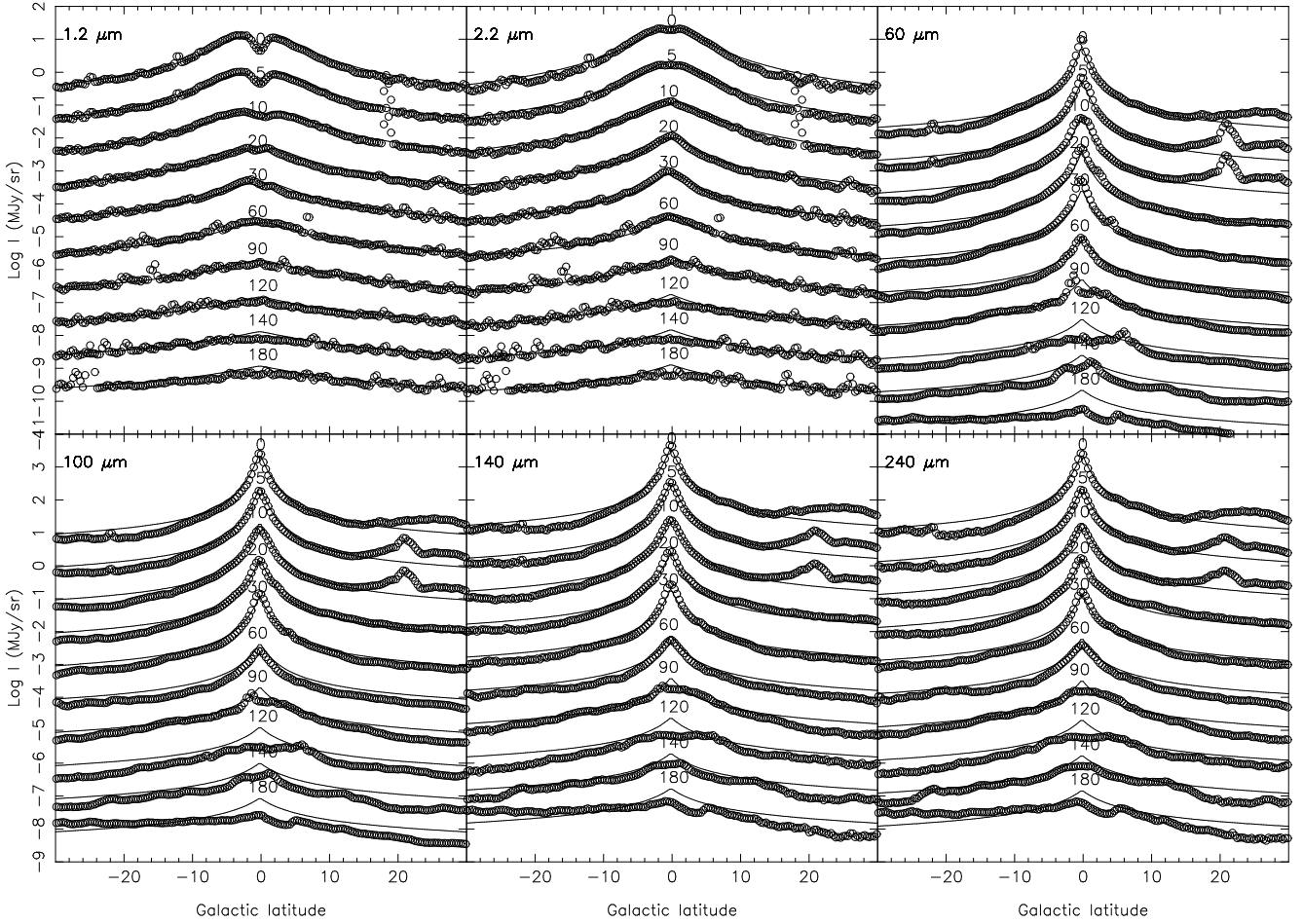}
\caption
{
Profiles along Galactic meridian zones (circles) together with the corresponding models (solid lines).
Taking advantage of the symmetry of the Galaxy, we have folded the maps over the central meridian
and extracted 10 meridian profiles (centered at 0, 5, 10, 20, 30, 60, 90, 120, 140, and 180 degrees in
Galactic longitude). Each profile is averaged over a 10 degree zone in Galactic longitude. Due to the
folding of the map, the zero longitude profile and the 180 longitude profile come from the 0-5 and 175-180
meridian zones, respectively. 
From top left to bottom right we show the profiles at 1.2, 2.2, 60, 100, 140, and 240$\mu$m (as indicated
in the upper left corner of each panel). 
}
\label{fig:verticuts}
\end{figure*}
%%%%%%%%%%%%%%%%%%%%%%%%%%%%%%%%%%%%%%%%%%%%%%%%%%%%%%%%%%%%%%%%

Having described the way to perform the fitting of the model to the DIRBE maps (Sect. \ref{sec:fit}),
we derive the values of the parameters of the model that best fit the data.
The fitted parameters are presented in Tables \ref{tab:firparams}, \ref{tab:nirparams},
and \ref{tab:gasparams}.
In Table \ref{tab:firparams} we give the values of the parameters that describe the dust
distribution inside the Galactic disk.
In Table \ref{tab:nirparams} we present the values of the parameters
for the stellar distribution as they have been determined from the 1.2 and 2.2 $\mu$m maps.
Finally, in Table \ref{tab:gasparams} we present the values of the parameters
for the distribution of the molecular and the atomic hydrogen.

Having determined all the parameters for the dust and the stellar distributions, we
can now create the model images of the Galaxy at each wavelength
to compare with the observed images.
This is what we do in Fig. \ref{fig:images}, where the observations at different wavelengths
(left panels) are compared with the model images (right panel).
The 1.2, 2.2, 60, 100, 140, and 240 $\mu$m images are shown  from top to bottom.
In these images the x- and y-axes are the Galactic longitude and latitude in degrees
with the Galactic center at (0,0), while positive and negative longitudes are the
southern and northern regions of the sky, respectively.
Positive and negative latitudes are regions above and below the Galactic plane, respectively.

As it is obvious from Fig. \ref{fig:images}, our model and the observations compare quite
well for all the different wavelengths.
The NIR (1.2, 2.2 $\mu$m) maps (two top panels) show the characteristic
dust lane crossing the bulge along the Galactic plane which, at these wavelengths, is poorly
visible (compared to the optical wavelengths). The stellar disk is well described by the
smooth exponential distribution used in the model (Sect. \ref{sec:model}). The FIR (60, 100, 140, and
240 $\mu$m) maps (third, fourth, fifth, and sixth panels from the top, respectively)
show the diffuse dust emission along the Galactic plane.
Given the simplicity of the axisymmetric distributions that were used, the model
quite accurately reproduces the observed emission at these wavelengths.

To get a better view of the goodness of the fit of the model to
the data, we produce vertical profiles of the surface brightness along 10
meridian zones of 10 degrees width each. This is shown in Fig. \ref{fig:verticuts},
where in each panel we overplot the model to the data for each wavelength (1.2, 2.2,
60, 100, 140, and 240 $\mu$m, as indicated with the wavelength value inside each
panel). In each panel (of different wavelength) the averaged profiles
along the longitude meridians of 0, 5, 10, 20, 30, 60, 90, 120, 140, and 180 degrees
are presented shifted with each other by a factor of 10 in surface brightness
for reasons of clarity. As one can see, the agreement between the model and the
observations is quite good.

% In all the plots the model is shown with a solid line and the data with
%open circles.

We note here that in the first two panels of Fig. \ref{fig:verticuts} the absorption by dust
is seen as a dip in the first few meridian profiles. The dip is more prominent at 1.2 $\mu$m
than at 2.2 $\mu$m. The 60 $\mu$m emission is mainly due to the warm dust that traces regions
of star formation. The 240 $\mu$m emission traces the diffuse cold dust.

To get a better feeling of the goodness of the fit,
we present the relative percentage of the residuals as a function
of the percentage of the Galaxy's area. We do that in Table 5 where,
for example, for the 1.2 micron band we have 29\% of the area
of the Galaxy with residuals less than 10\%, 53\% of the area
with residuals less than 20\%, and 92\% of the area of the
Galaxy with residuals less than 50\%. The same holds for the rest
of the bands. We see that, on average, about 50\% (half of the
Galaxy's area) has residuals less than 20\%. Given the large noise
measurements that exist in the high latitude regions (especially
in the FIR maps), and also the complexity of the real structures,
these numbers show how well the model fits the real data.

Using the values of the parameters presented in Table \ref{tab:firparams} and adopting
the Weingartner \& Draine (\cite{weingartner}) values for the extinction cross-section at
1.2$\mu$m and at 2.2$\mu$m, we can calculate the optical depth (and subsequently
the extinction) between any two
points in the Galaxy. In particular, the central face-on optical depth
$\tau_{\lambda}^f = 2 \kappa_{\lambda} z_c$
is 0.33 at 1.2 $\mu$m and 0.17 at 2.2 $\mu$m.

\begin{figure*}

\includegraphics[width=18cm]{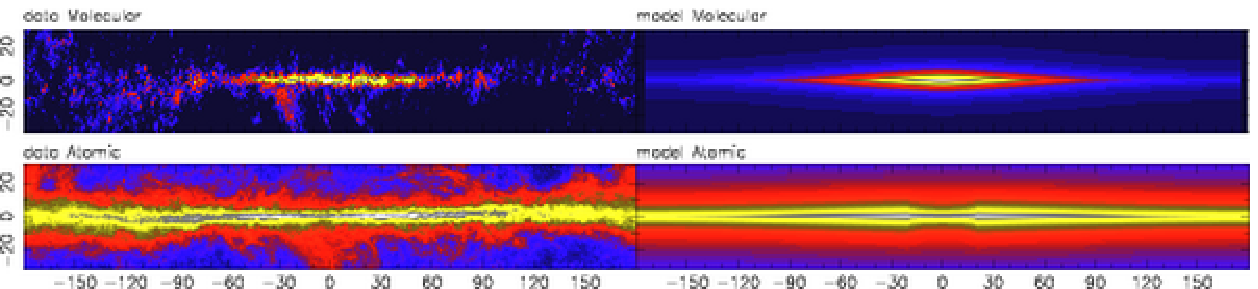}
\caption
{
The gas maps of the Galaxy (left panels) in direct comparison
with the fitted model (right panels). 
The upper left panel shows the molecular hydrogen distribution (Dame et al. \cite{dame}),
while the atomic hydrogen distribution (Kalberla et al. \cite{kalberla}) is presented in the
bottom left panel. 
Both maps are given for the Galactic latitude range of -30 to 30 degrees.
}
\label{fig:gasimages}
\end{figure*}

In Fig. \ref{fig:gasimages} we compare the maps of the hydrogen column
density with the corresponding model.
The top panel on the left shows the molecular hydrogen map as this is
inferred from the CO observations (Dame et al. \cite{dame}).
The respective model column density
is shown in the top right panel.
From the comparison between the model and
the observations it is evident that, despite the large clumpiness of the data, the large-scale
structure is fairly well represented by the model.
In the bottom left panel we show the map of the atomic hydrogen (Kalberla et al. \cite{kalberla}).
The lack of this component in the central part of the galaxy shows up as a decrease
of the column density.
The model column density is shown in the bottom right panel.
In this image, the lack of atomic hydrogen in the center of the Galaxy
is more prominent due to the sharp cut introduced by the model at the truncation radius $R_t$.

\begin{figure*}
\begin{center}
\includegraphics[width=12.1cm]{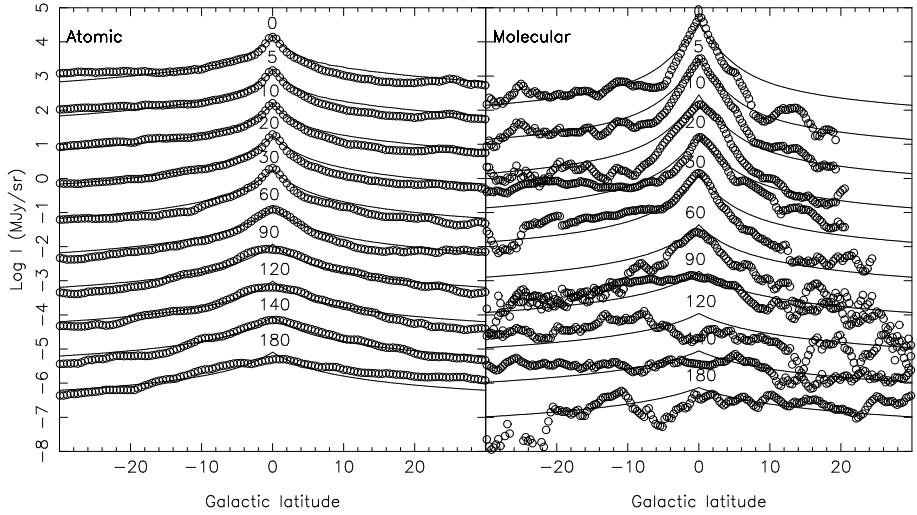}
\caption
{
Profiles along Galactic meridian zones (circles) together with the corresponding models (solid lines).
As in the case of the COBE/DIRBE data (Fig. 2), the maps are folded over the central meridian
and 10 meridian profiles are extracted following the same method described in Fig. \ref{fig:verticuts}.
The left panel shows the atomic hydrogen profiles, and the right panel shows the molecular hydrogen profiles.
}
\label{fig:gasverticuts}
\end{center}
\end{figure*}

The comparison between the model and the data is shown better in Fig. \ref{fig:gasverticuts}.
The left panel shows vertical profiles of the column density of the atomic hydrogen along
10 meridian zones of width 10 degrees each.
The agreement between the model and the data  is
good.
The observed CO (and subsequently the ${\rm H_2}$) distribution on the other hand
is very clumpy as it can be seen from the noise of the data in the profiles
shown in the right panel of Fig. \ref{fig:gasverticuts}.
The smooth model, however, follows the actual distribution
fairly well.

%%%%%%%%%%%%%%%%%%%%%%%%%%%%%% FIG 3 %%%%%%%%%%%%%%%%%%%%%%%%%%%
%%%%%%%%%%%%%%%%%%%%%%%%%%%%%%%%%%%%%%%%%%%%%%%%%%%%%%%%%%%%%%%%

%%%%%%%%%%%%%%%%%%%%%%%%%%%%%% FIG 4 %%%%%%%%%%%%%%%%%%%%%%%%%%%
%%%%%%%%%%%%%%%%%%%%%%%%%%%%%%%%%%%%%%%%%%%%%%%%%%%%%%%%%%%%%%%%

As in the case of the COBE/DIRBE maps, we also present the residuals
(in percentage) between the model and the observations as a function of the
Galaxy's area. We do that in Table 5 (last two rows).

Having determined the geometrical characteristics of the distributions and the 
central densities, it is then straightforward to compute the total mass of the cold dust
$M_c$, the warm dust $M_w$, the mass of the molecular hydrogen
$M_{\rm H_2}$, and the mass of the atomic hydrogen $M_{\rm HI}$ by integrating 
the respective three dimensional distributions in space. For distributions
of the form of Eqs. (1), (3), and (10) for the warm dust, the cold dust,
and the molecular hydrogen, respectively, this integration gives 
\begin{equation}
M=4\pi \rho(0,0) z h^2,
\end{equation}
with $h$ being the scalelength of the distribution, $z$ the scaleheight,
and $\rho(0,0)$ the central density of the material. If we also include
an inner cut-off in the distribution with a truncation radius $R_t$,
as we do for the atomic hydrogen (Eq. (11)), the integration then gives
\begin{equation}
M=4\pi \rho(0,0) z h (R_t+h) \exp \left( - \frac{R_t}{h} \right).
\end{equation}
Using the values for the fitted parameters presented in Tables \ref{tab:firparams} 
and \ref{tab:gasparams} and the above equations, we then find
$M_c=7.0\times 10^7 M_{\odot}$, 
$M_w=2.2\times 10^5 M_{\odot}$, 
$M_{\rm H_2}=1.3\times 10^9 M_{\odot}$, and 
$M_{\rm HI}=8.2\times 10^9 M_{\odot}$.
The total gas-to-dust mass ratio becomes $(M_{H_2} + M_{HI})/(M_c+M_w)=135$, which is in
fair agreement with the value of $160\pm60$ reported in Sodroski
et al. (\cite {sodroski}).

%%%%%%%%%%%%%%%%%%%%%%%% TABLE 2 %%%%%%%%%%%%%%%%%%%%%%%%%%%
\begin{table}
\centering
\caption[]{Parameters for the dust distribution (see text
for a detailed description of each parameter).}
\begin{tabular}{lll}
\hline
Parameter      &     Units  & Value                   \\
\hline
$\rho_{w}(0,0)$ & $\rm{gr~cm^{-3}}$ & $1.22\times10^{-27}$   \\
$h_w$           & $\rm{kpc}        $ & 3.3                   \\
$z_w$           & $\rm{kpc}        $ & 0.09                  \\
$\rho_{c}(0,0)$ & $\rm{gr~cm^{-3}}$ & $1.51\times10^{-25}$   \\
$h_c$           & $\rm{kpc}        $ & 5.0                   \\
$z_c$           & $\rm{kpc}        $ & 0.1                   \\
$T_c(0,0)$      & $\rm{K}     $ & 19.2                       \\
$h_T$           & $\rm{kpc}        $ & 48                    \\
$z_T$           & $\rm{kpc}        $ & 500                   \\
\hline
\end{tabular}
\label{tab:firparams}
\end{table}
%%%%%%%%%%%%%%%%%%%%%%%%%%%%%%%%%%%%%%%%%%%%%%%%%%%%%%%%%%%%%%%%
%%%%%%%%%%%%%%%%%%%%%%% TABLE 3 %%%%%%%%%%%%%%%%%%%%%%%%%%%%%%%%
\begin{table*}
\centering
\caption[]{Parameters for the stellar distribution (see text
for a detailed description of each parameter).}
\begin{tabular}{llll}
\hline
Parameter      &   Unit  & Value                   \\
\hline
               &          & $(\rm1.2\mu m)$  &  $(\rm2.2\mu m)$\\
\hline
$\eta_{disk}$ &$\rm erg \sec^{-1} cm^{-3} Hz^{-1} srad^{-1}$   &  $5.49\times10^{-38}$    & $9.94\times10^{-38}$       \\
$h_s        $ &$\rm kpc$                              &  2.5          & 2.2          \\
$z_s        $ &$\rm kpc$                              &  0.16         & 0.12         \\
$\eta_{bulge}$&$\rm erg \sec^{-1} cm^{-3} Hz^{-1} srad^{-1}$    &  $2.07\times10^{-34}$  & $2.03\times10^{-35}$  \\
$R_e$         &$\rm kpc$                              &  0.68         & 0.79         \\
$a/b$&--                                              &  0.61         & 0.63         \\
\hline
\end{tabular}
\label{tab:nirparams}
\end{table*}
%%%%%%%%%%%%%%%%%%%%%%%%%%%%%%%%%%%%%%%%%%%%%%%%%%%%%%%%%%%%%%%%%%%%

%%%%%%%%%%%%%%%%%%%%%%% TABLE 4 %%%%%%%%%%%%%%%%%%%%%%%%%%%%%%%%%%%%
\begin{table}
\centering
\caption[]{Parameters for gas distribution (see text
for a detailed description of each parameter).}
\begin{tabular}{lll}
\hline
Parameter      &   Units  & Value                   \\

\hline
$\rho_{H_2}(0,0)$        &$\rm cm^{-3}$      &  4.06      \\
$h_{H_2}        $        &$\rm kpc$                   &  2.57      \\
$z_{H_2}        $        &$\rm kpc$                   &  0.08     \\
$\rho_{HI}(0,0) $        &$\rm cm^{-3}$      &  0.32     \\
$h_{HI}        $        &$\rm kpc$                   &  18.24     \\
$z_{HI}        $        &$\rm kpc$                   &  0.52      \\
$R_t            $        &$\rm kpc$                   &  2.75      \\
\hline
\end{tabular}
\label{tab:gasparams}
\end{table}
%%%%%%%%%%%%%%%%%%%%%%%%%%%%%%%%%%%%%%%%%%%%%%%%%%%%%%%%%%%%%%%%%%%%

%%%%%%%%%%%%%%%%%%%%%%%%%%%%%% TABLE 5 %%%%%%%%%%%%%%%%%%%%%%%%%%%
\begin{table}
\centering
\caption[]{Residuals between the model and the observation.
For each COBE/DIRBE map we present the percentage of the 
Galaœô§Çy's area with residuals less than $<$10\% , $<$20\%, and $<$50\%.} 
\begin{tabular}{lcccc}
\hline
Wavelength ($\mu m$) & $<$10\%  & $<$20\% & $<$50\% \\
\hline
1.2                  &   29\% &  53\% &   92\% \\
2.2                  &   32\% &  55\% &   87\% \\
60                   &   26\% &  52\% &   88\% \\
100                  &   23\% &  43\% &   81\% \\
140                  &   28\% &  51\% &   81\% \\
240                  &   18\% &  37\% &   75\% \\
\hline
HI                   &   17\% &  38\% &   85\% \\
H$_2$                &   17\% &  35\% &   67\% \\
\hline
\end{tabular}
\label{tab:fitq}
\end{table}
%%%%%%%%%%%%%%%%%%%%%%%%%%%%%%%%%%%%%%%%%%%%%%%%%%%%%%%%%%%%%%%

%%%%%%%%%%%%%%%%%%%%%%%%%%%%%% FIG 5
\begin{figure*}
\resizebox{\hsize}{!}{\includegraphics{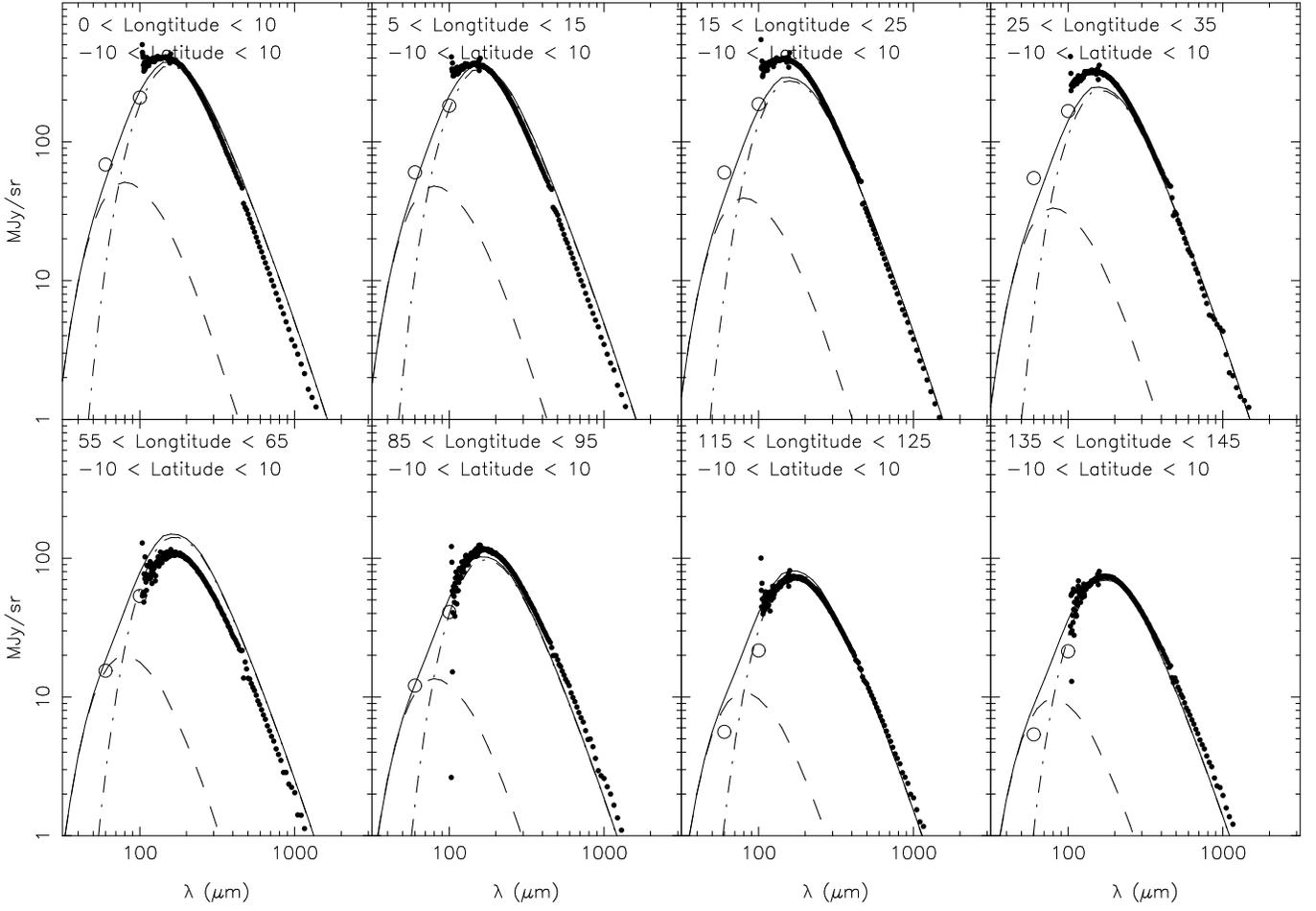}}
\caption
{
The Spectral Energy Distribution (SED) of the Milky Way averaged between -10 and 10 degrees in
Galactic latitude and over 10 degrees in Galactic longitude, as indicated on top of each panel.
In each panel the filled circles are the COBE/FIRAS data and the open circles are the COBE/DIRBE
data. The dashed line is the contribution to the SED of the warm dust component, the dot-dashed line
is the contribution of the cold dust component and the solid line is the total modeled SED.
}
\label{fig:firas}
\end{figure*}
%%%%%%%%%%%%%%%%%%%%%%%%%%%%%%%%%%%%%%%%%%%%%%%%%%%%%%%%%%%%%%%%%%%%%%%%%%%%%%%%%%%%%%%%%%%%%%%%

\section{Discussion}

\subsection{Comparison of the model with the FIRAS data}

To further establish the validity of the model, we compare it with
the COBE/FIRAS data. 
In Fig. \ref{fig:firas} we present the spectrum of the
Galaxy in several directions along the Galactic plane averaged over an
area of $\pm 10$ degrees in latitude and 10 degrees in longitude
centered on the Galactic equator and at 5, 10, 20, 30, 60, 90, 120, and 140
degrees in longitude. 
On average, the model is in good agreement with the data. The deviation
in some cases is to be expected due to the non-axisymmetric nature of
the real distribution of the dust. 

\begin{figure}
\includegraphics[width=8cm]{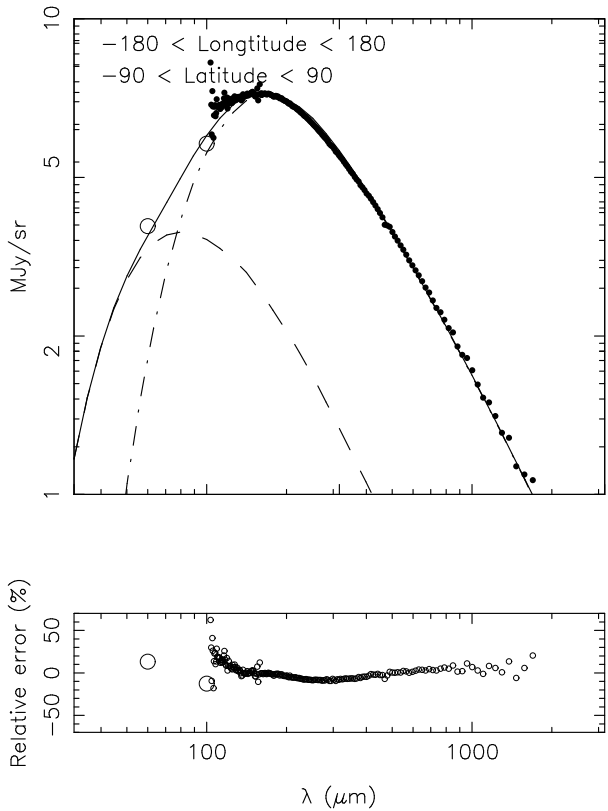}
\caption
{
The global SED of the Milky Way averaged over the whole sky (top panel).
As in Fig. \ref{fig:firas}, the filled circles are the COBE/FIRAS data and the open circles are the COBE/DIRBE
data, while the dashed line is the contribution to the SED of the warm dust component, the dot-dashed line
is the contribution of the cold dust component, and the solid line is the total modeled SED.
In the bottom panel we present the residuals (in percentage) between model and real measurements. 
}
\label{fig:allfiras}
\end{figure}

Evidence supporting the previous statement comes from a comparison of
the ``global'' SED of the Milky Way constructed by averaging the
COBE/FIRAS maps in all the directions on the sky with the predicted
SED from our model. This is shown in Fig. \ref{fig:allfiras} (top panel). 

The goodness of the fit can be evaluated by looking at the 
residuals between the model and the observations which are presented in
the bottom panel of Fig. \ref{fig:allfiras}. From this plot we
see that 82\% of the data points on this SED show residuals from
the modeled SED that are less than 10\%. In particular, 14\%, 24\%, 47\%, 82\%,
and 93\% of the data points show residuals from the model which are less than
1\%, 2\%, 5\%, 10\%, and 20\%, respectively. 

This agreement, which does not come from fitting the model to the data, but
from comparison of the data with our model, gives us confidence
that our model gives a consistent representation of the Galaxy in
wavelengths extending up to 1000 $\mu$m.

In their analysis, Reach et al. (\cite{reach}) reported emission from
a very cold component (4-7 K). This component is not evident in our
analysis when looking at the averaged SED of the Galaxy
(Fig. \ref{fig:allfiras}). However, deviations though between the model and
the observations do exist when looking through various lines-of-sight
(Fig. \ref{fig:firas}), which can, as said earlier, be explained by
local deviations of the density of the dust material.  

%%%%%%%%%%%%%%%%%%%%%%%%%%%%%% FIG 6 %%%%%%%%%%%%%%%%%%%%%%%%%%%%%%%%%
%%%%%%%%%%%%%%%%%%%%%%%%%%%%%%%%%%%%%%%%%%%%%%%%%%%%%%%%%%%%%%%%%%%%%%%

\subsection{Temperature distribution}

%%%%%%%%%%%%%%%%%%%%%%%%%% FIG temp %%%%%%%%%%%%%%%%%%%%%%%%%%%%%%%%%
\begin{figure*}
\includegraphics[width=18cm]{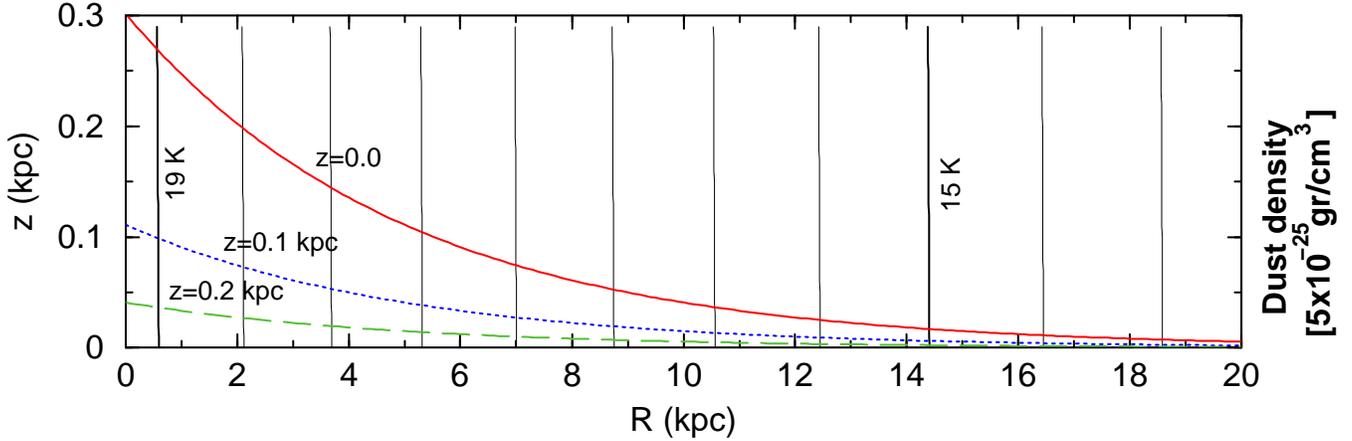}
\caption
{Temperature map of the cold dust component within the Milky Way.
Temperature contours (the almost vertical black solid lines) are
plotted every 0.5 K and highlighted at 15 K  and 19 K by a label
and a thicker line. The radial profiles of the cold dust density
are also overplotted for z=0.0 (red solid line), z=0.1 kpc (blue
dotted line), and z=0.2 kpc (green dashed line) in units of
$5\times 10^{-25}$ gr/cm$^3$.}
\label{fig:temp}
\end{figure*}
%%%%%%%%%%%%%%%%%%%%%%%%%%%%%%%%%%%%%%%%%%%%%%%%%%%%%%%%%%%%%%%%%%%

As already  mentioned in Sect. 3.1, the dust in our model is described by two components
(one with a constant ``warm'' temperature of 35 K, associated with 
the warm HII star-forming regions, and another
``cold'' temperature component). For the distribution of the temperature 
of the cold dust material, we assume a simple exponential distribution
in both directions, radially and vertically to the plane of
the Galactic disk (see Eq. \ref{eq:colddusttemp}), with a boundary condition 
that the temperature at large distances is at 3 K (of course, as noted earlier,
the radiative transfer calculations only take place within a finite
volume where dust material exists; see Sect. 3). To better
visualize the spatial distribution of the temperature of the cold dust
component we plot the temperature map seen in Fig. \ref{fig:temp}.
This map represents a two dimensional slice of the Galactic plane that
extends up to 20 kpc along and up to 0.3 kpc above the Galactic plane
with the temperature contours plotted every 0.5 K. 
The radial profiles of the cold dust density 
(in units of $5\times 10^{-25}$ gr/cm$^3$)
are also overplotted for z=0.0, z=0.1 kpc, and z=0.2 kpc indicating the
way that the dust material is distributed within this temperature
field.
From this plot we see that the temperature varies from 19.2 K 
at the center to
$\sim 15$ K at the outer parts of the Galaxy ($\sim 15$ kpc).
This result agrees with the studies of Reach et al. 
(\cite{reach}), Sodrosky et al. (1994), and Davies et al. (1997)
showing the similar radial dependence of the cold dust component.
Concerning the vertical dependence of the temperature of the
cold component, we see that this is more or less constant for
different heights above the Galactic plane. We note here though 
that for higher opacities this picture could be different, with 
the dust being more effectively heated in
places above the disk compared with the dust in the Galactic
plane, where the dust is shielded from the diffuse radiation field
(see Bianchi et al. 2000).

\subsection{Star formation efficiency in the Milky Way disk and the Schmidt law}
As already mentioned in the Introduction, a good indicator of the SFR is the
FIR emission (e.g., Hunter et al. 1986; Lehnert \& Heckman 1996;
Meurer et al. 1997; Kennicuut 1998b; Kewley et al. 2002). 
To exploit the FIR emission we make use of
the strong correlation between the SFR and the 100 $\mu$m flux found in
Misiriotis et al. (\cite{misiriotis_2004}) for a large sample of spiral galaxies. 
Using this relation and calculating the total 100 $\mu$m luminosity, 
we derive a SFR of ${\rm 2.7 M_{\odot}yr^{-1}}$ for the Milky Way. 

\begin{figure}
\resizebox{\hsize}{!}{\includegraphics{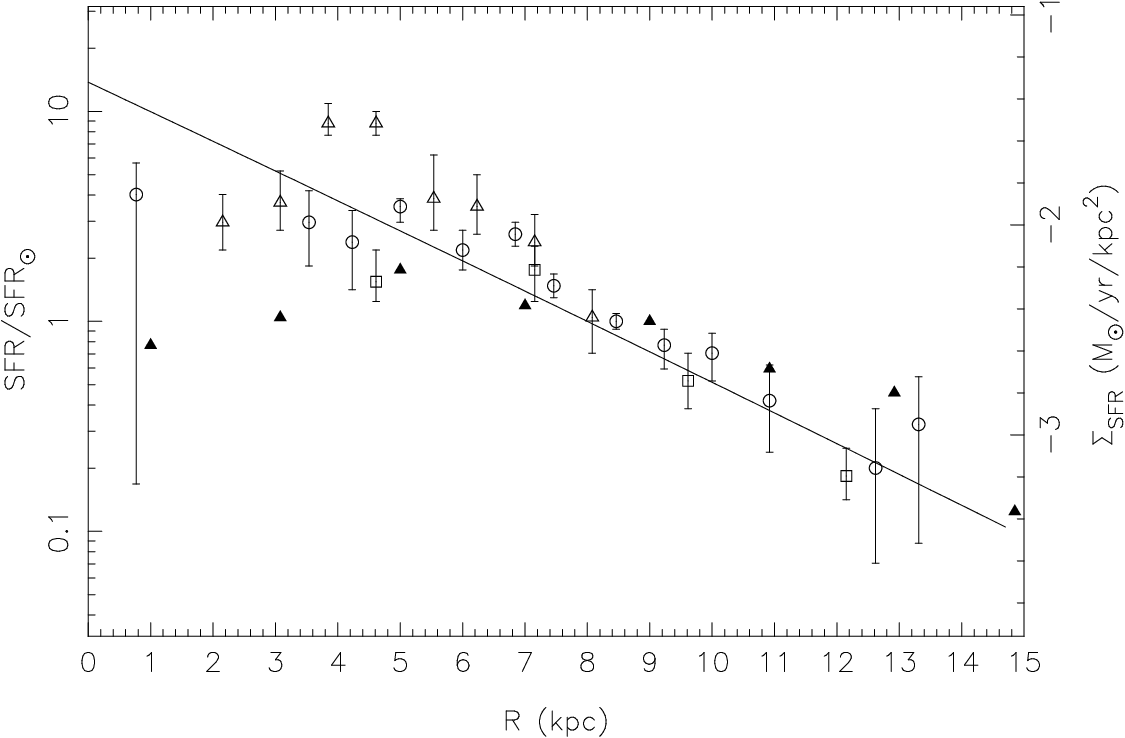}}
\caption
{
The star formation rate (SFR) normalized to the value at the galactocentric
distance of 8 kpc (SFR$_\odot$; y-axis on the left) and the star formation rate
density ($\Sigma_{SFR}$; y-axis on the right) as a function of radius in the Milky Way.
The data correspond to different star formation estimates and are taken from
Lyne et al. (\cite{lyne}, open circles), Guesten \& Mezger (\cite{guesden}, triangles), and
Guibert et al. (\cite{guibert}, squares; see Fig. 2 of Boissier \& Prantzos \cite{boissier}).
The solid line corresponds to the model prediction.
}
\label{fig:sfrradial}
\end{figure}

The SFR surface density can then be calculated in the disk of the Milky Way,
using the same relation presented in Misiriotis et al. (\cite{misiriotis_2004}), 
and plotted as a function of the Galactic radius. This is done in Fig. \ref{fig:sfrradial}, 
where our model prediction (solid line) is compared with several estimates 
of the SFR found in the literature (see Boissier \& Pratzos \cite{boissier} and 
references therein). Both the SFR normalized to the value at the
galactocentric distance of 8 kpc (SFR$_{\odot}$) and the star formation
rate density are presented in this plot as a function of R, with the
model showing an excellent agreement with the existing estimates of
the various SFR tracers to at least two kpc from the center. The inner region
shows, acknowledging the large uncertainty of the few data points available, a decrease of 
the SFR efficiency, presumably due to a non-exponential distribution of
the dust in such small scales. Lack of dust, for example, in this region may be due
to the presence of the bar structure of the Galaxy. As mentioned earlier 
though, the purpose of this study is to keep the modeling as simple as 
possible with the aim of deriving a valuable description of the Galaxy. 
With this goal in mind we avoid going into a more sophisticated modeling of the central part.

A simple power-law relation between SFR and the gas content of external galaxies,
introduced by Schmidt (\cite{schmidt}) and further explored by Kennicutt (\cite{kennicutt_1998a}), 
is well established and tested for large samples of galaxies (Misiriotis et al. \cite{misiriotis_2004}). 
This relation is expressed in terms of the SFR 
surface density $\Sigma_{SFR}$ and the gas surface density $\Sigma_{\rm gas}$  
and has the form:
\begin{equation}
\Sigma_{SFR}=A\Sigma_{\rm gas}^N,
\end{equation}
with $A$ being the absolute SFR efficiency and $N$ the slope of the
power-law. 
Kennicutt (\cite{kennicutt_1998a}) examined two samples of galaxies showing normal and starbursting
behavior and derived a Schmidt law for normal galaxies (with $N=2.47\pm0.39$) and
for starburst galaxies (with $N=1.40\pm0.13$). 

\begin{figure}
\resizebox{\hsize}{!}{\includegraphics{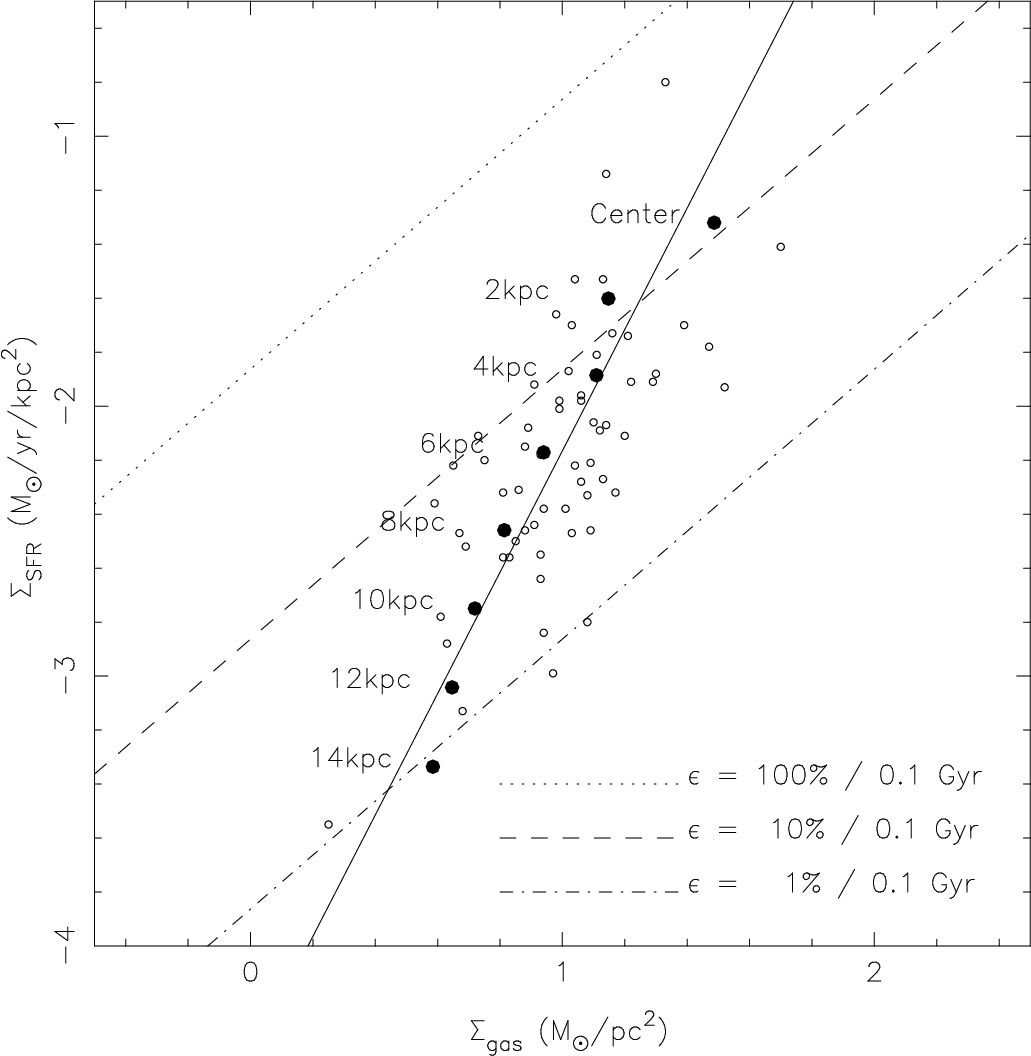}}
\caption
{
Star formation rate density as a function of gas surface density
in the Milky Way disk. The solid circles indicate the measurements
for the Milky Way (for distances along the disk as indicated by the
numbers close to the solid circles), while the solid line is the
best fit of a Schmidt law to these points (see the text for details).
The open circles are the global measurements for a sample of
external galaxies presented in Kennicutt (\cite{kennicutt_1998a}).
The three  parallel dashed and dotted lines correspond to constant star formation
efficiencies of 1\%, 10\%, and 100\% per 0.1 Gyr.
}
\label{fig:sfrgasdensity}
\end{figure}

In our study we have,
for the first time, the opportunity to examine the validity of
this law not for a sample of external galaxies, by measuring 
their global properties, but for different
regions of the same galaxy (the Milky Way). This is shown in 
Fig. \ref{fig:sfrgasdensity} with the SFR surface density ($\Sigma_{SFR}$) calculated for different 
radii along the Galactic disk, plotted against the gas surface density
$\Sigma_{\rm gas}$ of the same region. These data are shown
with the numbers close to the points indicating the distance to the 
Galactic center (in kpc).
Furthermore, it is evident that for regions within the Galaxy with small 
gas content (at the outer part of the disk) the SFR is low,
while it gets more intense for regions with
larger gas density (close to the center of the galaxy).
From this plot it is evident that the Galactic Schmidt law also follows a
power law. A best fit to these data (indicated with a solid line in the plot) yields
\begin{equation}
{\rm
\frac{\Sigma_{SFR}}{M_{\odot}yr^{-1}pc^{-2}} = 
4.45 \times 10^{-5} \left(\frac{\Sigma_{gas}}{M_{\odot}kpc^{-2}}\right)^{2.18}.
}
\end{equation}
The power-law index of $2.18\pm0.20$ that we find is in a very good agreement with the value
of $2.47\pm0.39$ quoted by Kennicutt (\cite{kennicutt_1998a}) for normal galaxies. This is also
evident by looking at the distribution of the normal galaxies  
on this plot (Fig. \ref{fig:sfrgasdensity}). The points for the external galaxies 
are well correlated with the measurements for different regions within the Milky Way. 
This indicates the strong correlation between the gas content
and the SFR activity. The striking similarity of the ``internal'' Schmidt law
for an individual galaxy (in this case the Milky Way) and the ``external'' Schmidt
law for the global properties of normal galaxies may indicate a universal gas-SFR
correlation not only on a large scale (for galaxies as a whole), but also on a 
small scale (for different regions inside a galaxy). 

%%%%%%%%%%%%%%%%%%%%%%%% FIG 7 %%%%%%%%%%%%%%%%%%%%%%%%%%%%%%%%
%%%%%%%%%%%%%%%%%%%%%%%%%%%%%%%%%%%%%%%%%%%%%%%%%%%%%%%%%%%%%%%%%%%%%%
%%%%%%%%%%%%%%%%%%%%%%%% FIG 8 %%%%%%%%%%%%%%%%%%%%%%%%
%%%%%%%%%%%%%%%%%%%%%%%%%%%%%%%%%%%%%%%%%%%%%%%%%%%%%%%%%%%%%%%%%%%%%%%%

Another very useful parameter that describes the current star-forming activity 
of a galaxy is the star formation efficiency ($\epsilon$). In Fig. \ref{fig:sfrgasdensity}
the three parallel, lines correspond
to constant SFRs per unit gas mass in units of 1\%, 10\%, and 100\% per 0.1 Gyr,
as described in Kennicutt (\cite{kennicutt_1998a}). It is interesting to notice that the star formation
efficiency of the Milky Way goes from $\sim$1\% per 0.1 Gyr in the outer regions ($\sim 14$ kpc)
to $\sim$10\% per 0.1 Gyr in the center. This means that the Galaxy converts $\sim$1\%
of the gas in the outer parts of the disk ($\sim$10\% in the center) to stars over
the period of 0.1 Gyr, which roughly corresponds to one orbital period of the disk.

The average global efficiency of star formation in the Milky Way can be
calculated by dividing the SFR over the period of 0.1 Gyr ($2.7\times 10^8 M_{\odot}$;
see beginning of Sect. 6.3) by the total gas mass ($M_{H_2} + M_{HI} = 9.5\times 10^9$).
This results in 2.8\% per 0.1 Gyr or, in other words, that our Galaxy
spends 2.8\% of its gas to create stars over 0.1 Gyr. This
is at the lower end of the median star formation efficiency of
typical present-day spiral galaxies, which is 4.8\% (Kennicutt \cite{kennicutt_1998a}).

\subsection{Comparing the model with optical data}
\begin{figure}
\resizebox{\hsize}{!}{\includegraphics[angle=-90.]{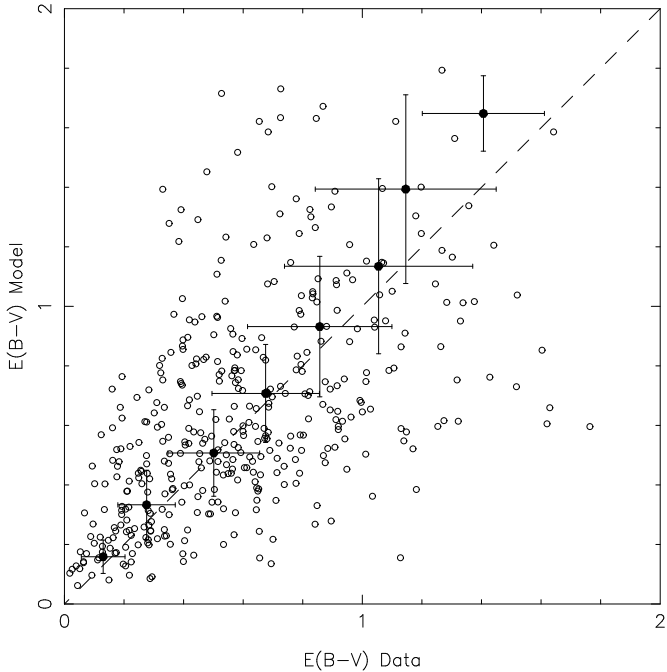}}
\caption
{The B-V color excess for the sample of the Galactic Classical Cepheids
(Fernie et el. 1995; x-axis) compared with the color excess along the
same directions as derived from the model.}
\label{fig:cepheids}
\end{figure}

To check the validity of our model in the optical wavelengths,
we use the Database of Galactic Classical Cepheids
(Fernie et el. \cite{fernie}). This database contains information about over 500
classical Cepheids in the Galaxy with determined distances ranging from a
couple of hundred parsec up to $\sim 6.5$ kpc. The E(B-V) color
excess reported in this catalog for each star was then checked against the
E(B-V) determined with our model by calculating for each Cepheid the extinction in the B- and V-bands.
The B and V extinction maps for the Galaxy were created by assuming 
the values of $h_c$ and $z_c$ given in Table 2 and using the
values of the extinction coefficient for the B- and V-bands calculated in 
Weingartner \& Draine (\cite{weingartner}).
The comparison between the observed data and the model predictions for E(B-V)
is shown in Fig. \ref{fig:cepheids} with the x-axis being the B-V color excess observed and the
y-axis the B-V color excess predicted by the model. From this plot it is evident that
the correlation between the observed and the predicted measurements 
is extremely good with the average of the points, binned every
0.2 magnitudes of E(B-V) on both axes, falling, within the error bars, along the
diagonal of the plot. This is also evident by looking at the individual observations
that are scattered symmetrically with respect to the diagonal.

The scatter of the points indicates that in many cases our model either overestimates
or underestimates the amount of dust along the line of sight. This is clear evidence of
the clumpy distribution of the dust. However, the fact that the opacity overestimated lines
of sight are about as many as the underestimated ones ensures that on average our smooth
model is consistent with the observed extinctions.

%%%%%%%%%%%%%%%%%%%%%%%% FIG 9 %%%%%%%%%%%%%%%%%%%%%%%%%%%
%%%%%%%%%%%%%%%%%%%%%%%%%%%%%%%%%%%%%%%%%%%%%%%%%%%%%%%%%%%%%

\section{Conclusions}
In the present study a consistent model
of the ISM for a wide range of the spectrum (from optical wavelengths to
millimeter wavelengths) is presented. 
Using a simple axisymmetric three-dimensional
radiative transfer model for the dust and the stars in the Galactic
disk, we derive the set of the parameters that best describes the FIR emission
observed by COBE/DIRBE at 60, 100, 140, and 240 $\mu$m. Having constrained
the dust distribution, we then fit the NIR COBE/DIRBE observations at 1.2 and 2.2 $\mu$m
and determine the NIR properties of the Galaxy. 
The scalelength for the stars at 1.2 and 2.2 $\mu$m is found to be
2.5 and 2.2 kpc respectively. This comes in agreement with findings
of recent studies (see Drimmel \& Spergel 2001 and references therein).
Having calculated both the NIR bands, we see the trend of increasing
scalelength with decreasing wavelength. This trend is also found in
other sudies of nearby galaxies (e.g., De Jong 1996; Xilouris et al. 1999).
We find that the ratio of the stellar to the dust scaleheight is $\sim 1.7$,
while the scalelength of the dust is about twice that of the stars
(as inferred from the NIR observations). Given the fact that we expect 
larger scalelength values in the optical (by as much as 20\%), this 
scalelength ratio could become $\sim 1.4 - 1.5$ in the optical
wavelengths (see Davies et al. 1997; Xilouris et al. 1999).
In addition to the COBE data, we also model the 3D distribution of the
atomic and molecular hydrogen as they reveal themselves from the
21 cm and the CO emissions, respectively.
 We then compare our model predictions
(for wavelengths up to 1000 $\mu$m) with the observations by COBE/FIRAS. 
The global SED  of the Galaxy (averaged over the entire Galaxy) is in excellent agreement
with the model, while the model shows small deviations from the SED of 
specific directions on the sky. These could be explained by assuming local
deviations of the density of the dust (clumpiness of the dust).
The star formation rate as a function of Galactocentric distance is then
derived by the model and is found to be in excellent agreement with various star formation 
tracers within the Galaxy. Comparing the SFR surface density with the gas
surface density, an ``internal'' Schmidt law is derived for the regions 
along the Galactic disk that is almost identical to the ``external'' Schmidt
law derived by Kennicutt (\cite{kennicutt_1998a}) for typical spiral galaxies. This could 
be an indication of a universal gas-SFR behavior not only on large scales
(for the galaxies as a whole), but also on small scales (within an individual 
galaxy). The star formation efficiency of the Milky Way is found to decrease
from $10\%$ to $1\%$ per 0.1 Gyr when going from the center to the outer parts of the Galactic
disk. Finally the model predictions in the optical wavelengths (B- and V-bands)
are compared with existing observations of Cepheid stars (in terms of their
observed E(B-V) color excess) and a good correlation (within the statistical 
errors) is found. 

\begin{acknowledgements}
We thank N. D. Kylafis and S. Bianchi for their significant contribution in the final stages of this work.
We would also like to thank V. Charmandaris for useful discussions
concerning the FIRAS data and P. M. W. Kalberla for kindly providing us with the map of the
Leiden/Argentine/Bonn Survey of Galactic HI.
We acknowledge the use of the Legacy Archive for Microwave Background Data Analysis (LAMBDA). 
Support for LAMBDA is provided by the NASA Office of Space Science.
We would like to thank the anonymous referee for pointing several
weaknesses in the early version of this paper.
This work has been supported in part by a Pythagoras II research program of the
Ministry of Education of Greece.
\end{acknowledgements}

\end{document}